# Pedagogical techniques of Earth remote sensing data application into modern school practice

Ihor V. Kholoshyn[1\[0000-0002-9571-1758\]], Iryna M. Varfolomyeyeva[1\[0000-0002-0595-524X\]],
Olena V. Hanchuk[1\[0000-0002-3866-1133\]], Olga V. Bondarenko[1\[0000-0003-2356-2674\]]
and Andrey V. Pikilnyak[2\[0000-0003-0898-4756\]]

[1] Kryvyi Rih State Pedagogical University, 54, Gagarina Ave., Kryvyi Rih, 50086, Ukraine
`holoshyn@kdpu.edu.ua`, `iravarfolomeeva365@gmail.com`,
`elena.ganchuk@gmail.com`, `bondarenko.olga@kdpu.edu.ua`
[2] Kryvyi Rih National University, 11, Vitali Matusevich Str., Kryvyi Rih, 50027, Ukraine
`pikilnyak@gmail.com`

**Abstract.** The article dwells upon the Earth remote sensing data as one of the basic directions of Geo-Information Science, a unique source of information on processes and phenomena occurring in almost all spheres of the Earth geographic shell (atmosphere, hydrosphere, lithosphere, etc.).

The authors argue that the use of aerospace images by means of the information and communication technologies involvement in the learning process allows not only to increase the information context value of learning, but also contributes to the formation of students' cognitive interest in such disciplines as geography, biology, history, physics, computer science, etc.

It has been grounded that remote sensing data form students' spatial, temporal and qualitative concepts, sensory support for the perception, knowledge and explanation of the specifics of objects and phenomena of geographical reality, which, in its turn, provides an increase in the level of educational achievements.

The techniques of aerospace images application into the modern school practice have been analyzed and illustrated in the examples: from using them as visual aids, to realization of practical and research orientation of training on the basis of remote sensing data.

Particular attention is paid to the practical component of the Earth remote sensing implementation into the modern school practice with the help of information and communication technologies.

**Keywords:** Earth remote sensing data, aerospace images, Geo-Information Science techniques.

## 1 Introduction

### 1.1 Scientific relevance of the research

Earth remote sensing data (RS) is a unique source of information about the processes occurring in the Earth's geographic shell, and therefore their role in the study of geography is constantly increasing. Aerospace images are more effective than the terrestrial information system in terms of their information, since they enable to obtain information with the required spatial-temporal resolution and the image of the Earth's surface in spectral ranges of various radiations.

Aerospace images are of undoubted interest for the educational process. Using remote sensing data, one can visualize the natural and anthropogenic processes and evaluate their dynamics; inform about the placement of objects on the Earth's surface; to monitor (constant monitoring) the state of geographic objects and processes, to formulate hypotheses or to identify patterns.

The unusual character and novelty of such information is of great interest in terms of new technologies, and as a consequence, and of a more in-depth study of basic educational subjects. According to Svetlana S. Karimova and Michael B. Veselov [8], the main educational advantages of remote sensing data are: a large degree of visibility, which is generally unattainable for traditional geographic maps; high resolution; high realism; reflection of their objective reality; great depth of consideration of the investigated object or phenomenon; increasing the possibilities of demonstrating the complexity and interconnection of the processes; increased attention to the physical foundations of the studied processes, etc.

The use of aerospace images in the school geography course began in the early 1950s, when the results of aerial photography appeared in the Soviet school atlases and textbooks. However, wide-ranging use of geospatial data did not receive the results of remote sensing, due to the lack of methods for using Earth images from space in the learning process, as well as the closure of most remote-sensing materials for open use.

Since the mid-80s of the last century, theoretical and practical researches based on modern technologies have been actively conducted abroad, using remote sensing data as educational resources.

At the beginning of the third millennium, qualitative changes in the use of remote sensing data for educational purposes began to take place. Openness and accessibility of information, active computerization of the educational process and the widespread use of Internet technologies contributed to the formation of a new pedagogical direction – space geography. Space geography makes a special emphasis on the formation of the most complete and close to the true reality of the visual image of various geographic objects through the study of their portrait space models. In the process of decoding these models, an understanding of the inter-component natural connections, the economy and the population with the natural environment is fixed [3; 10; 13; 16; 17].

The main mechanism for the introduction of satellite imagery in school practice is the various educational programs initiated by large space companies and organizations (NASA, OSC, Rocketdyne, etc.). Their essence is to encourage the pedagogues of some certain educational institutions to integrate the application of remote sensing data into the school curriculum. For this purpose, educational establishments are provided with the necessary programs and software, computer equipment and receivers of satellite signals.

**1.2 Recent research and publications analysis**

A significant contribution to the development of the theory and methodology of aerospace images use as an educational resource belongs to such scholars as: Alexander V. Barladin [1] (preparation of remote sensing data for using in multimedia presentations), Alexander M. Berlyant [2] theoretical foundations of geoinformatics), David Richard Green [7] (using GIS technologies at school), Lyudmila M. Datsenko and Vitaliy I. Ostroukh [4] (studying the foundations of geoinformation systems and technologies in special-field profile education), Nakis Z. Khasanshina [9] (the potential of geo-informational technologies in teaching geography), Ihor V. Kholoshin [11] (pedagogical techniques of Earth remote sensing data application into modern school practice), Rimma D. Kulibekova [12] (geoinformation technology as a means of information culture formation with the future geography teacher), Witold Lenart, Anna Wozniak, Malgorzata Witecka [19] (GIS at school), Vladimir S. Morkun, Serhiy O. Semerikov and Svitlana M. Hryshchenko [14] (methods of using geoinformation technologies in mining engineers' training), S. Simone Naumann, Alexander Siegmund, Raimund Ditter and Michelle Haspel [15] (theory and practice of the Earth's remote sensing), Oleh M. Topuzov [18] (informatization of geographic education) and others.

However, the question in what form, with the use of information and communication technologies and methodical techniques, the use of remote sensing in the teaching of the school geography course is possible, remains open.

**1.3 Article objective**

The objective of the proposed study is to analyze pedagogical techniques for the introduction of Earth remote sensing data into the practice of a modern school using ICT.

**2 Research results**

Aerospace images have all the necessary features that are characteristic of geography training [6]. Let us describe some of them. First, they contain training information that allows them to be used as a source of knowledge, and secondly, they can be used during practical work to develop skills and abilities.

The information obtained in the study of aerospace images determines the specifics of their use in the learning process and opens new opportunities for remote sensing data as educational resources (Table 1). For example, the opportunity to see how geographic objects look in real form from a height in the range of 100 meters to tens of thousands of kilometers significantly increases the visibility of learning, making it more figurative, bright and memorable.

The complex nature of the information that is read from aerospace images (relief, fauna and flora, meteorological factors, socio-economic aspects) provides a comprehensive approach to information analysis, through the acquisition of knowledge on related disciplines: biology, geology, medicine, ecology and etc. As a result, the students master such methods as analysis and synthesis, make logical generalization and conclusions that significantly activate the student's creative activity, increase their motivation to acquire new knowledge.

**Table 1.** Patterns of students' skills formation and pedagogical result due to the remote sensing data characteristics

| Earth remote sensing data characteristics | Skills formation | Pedagogical result |
|---|---|---|
| Real image of the objects being studied | Formation of the investigated object (phenomenon) image on the basis of decoding aerospace images | It makes learning more figurative, bright and memorable |
| Complex character of information read from aerospace images | Mastering the methods of analysis and synthesis, the ability to build logical inferences and draw conclusions. | It activates the student's creative activity, increases motivation to acquire new knowledge |
| Monitoring the territory in time and space | Analysis of spatial and temporal information, modeling and predicting situations | It develops the potential of students' cognitive activity, involves them in research work |
| Great practical value of the information obtained when decoding aerospace images | Assessment of the studied areas state (accounting for the dynamics of changes in natural and anthropogenic factors) | It strengthens the influence that brings up training, forms practical skills and an active life position |

One of the advantages of remote sensing data is the ability to monitor the territories for a long time, to provide the learning process with the sources of knowledge and the means necessary to carry out practical training. Based on these data, students are given the opportunity to study objects and phenomena in space and time, modeling and predicting the situation. It develops the potential to cognitive activity, involves them in research work.

Remote sensing data is a source of unique information of great practical significance. By its very nature, the aerospace image is a spatial model that replaces real objects and phenomena. At the same time, the picture performs a dual role: it is a means of research, on the one hand, and an object of research, on the other. A detailed study of the images helps to form an objective holistic image of the studied areas with their spatial-temporal characteristics, which is necessary for a comprehensive assessment of their state. Realizing the reality by analyzing airborne images in the process of studying geography proceeds in several stages [11] (Fig. 1).

The first stage – the understanding of aerospace images, involves the formation of knowledge among students about the main characteristics of images, the main features of reflecting various geographical objects, processes and phenomena on them. In fact, at this stage, the students lay the foundation for the practical use of remote sensing data.

The second stage – images decoding, is the ability to distinguish and recognize geographical objects (phenomena), as well as to identify their qualitative and quantitative indicators. This is the main focus in realizing reality through aerospace imagery, because it is at this stage that students learn about the basic characteristics of reality.

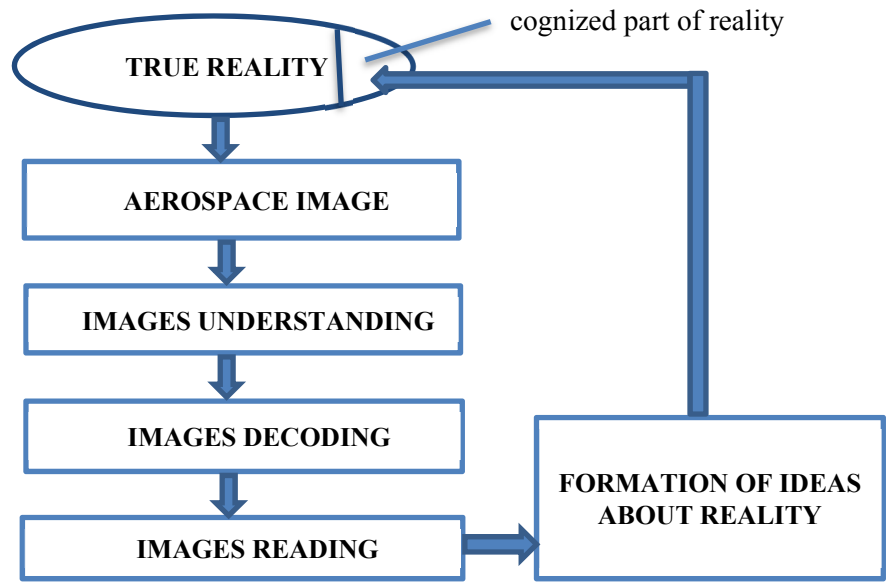

**Fig. 1.** Scheme of reality cognition by studying aerospace images in the process of geography training

The third stage – the reading of aerospace images involves mastering the means of compiling descriptions of geographical objects and phenomena based on the results of decoding Earth images from outer space. At this stage, the students form the basis of images decoding in terms of their location, state, interconnection and dynamics of real objects and phenomena. Creating an image, analyzing and interpreting it using inductive and deductive generalizations leads to the expansion and enrichment of knowledge about the investigated reality, which promotes the formation of practical skills and active life position. At the same time, the teacher must be able to explain to students that the image created by them may differ significantly from reality, since the picture only conveys a part of it.

In the learning process, Earth remote sensing can perform various functions.

## 2.1 Earth remote sensing data as a means of visual aids

The main visual guide when studying geography is the map. Given its advantages, it should be noted at the same time that the cartographic representation of objects is very arbitrary and does not accurately reflect the objective reality. Aerial imagery forms students' visual image of the objects and phenomena studied, which contributes to a more specific perception of their essence and a more qualitative memory of the of the educational material contents.

In the process of geographical representations visual formation, the means of remote sensing can be divided into four stages:

1. visual representation of definite geographical objects and phenomena characteristics;
2. occurrence of representations about geographical objects and phenomena;
3. preservation and reproduction of representations about geographical objects, phenomena or their certain characteristics;
4. regular application of the received ideas into the process of concepts formation.

It is possible to apply several ways of presenting remote sensing data using information and communication technologies. The most convenient, expedient and obsolete form of aerospace images is their presentation on various digital media. Low cost, saving a large amount of information, and most importantly, high visual characteristics, contribute to the dominance of this form of data representation of remote sensing data. The image is displayed on the computer monitor screen, although the use of the computer-plus-projection device demonstration is particularly effective. Designing aerospace images on a large screen greatly enhances the visibility of learning, as well as increases motivation to mastering it.

It is important that the teacher uses aerospace images not only as static information, but also creates the foundation for mental operations. Thus, using multi-dimensional space images, historical or geological materials, students can be taught elements of geographic modeling and predicting.

Aerospace images make it possible to grasp the role of geography as an actual contemporary science, which, while studying the environment, can significantly affect the development of many aspects of human activity and the interaction of man with nature. The teacher by specific examples (Aral Photos, Gulf Coast catastrophe, etc.) demonstrates how remote sensing data can monitor the anthropogenic impact on nature.

Aerospace images as visual aids can be used at different stages of the lesson for different purposes. The most typical is the use of shots when explaining the learning material by a teacher to form a geographic image of objects and phenomena. It is also advisable to use the remote sensing data before studying the topic. In this case, they will act as a means of forming initial ideas and motivating cognitive activity. In addition, photos can be used to fix the training material and to control the data received.

Fig. 2 shows an example of the MyTest program, which is a consortium of electronic tests. As a means of visualizing test questions, aerospace images are used.

With MyTest (and the like, e.g., Test Designer), you can organize and test students' knowledge of both single topics and entire school geography courses. No less effective is the use of the program for educational purposes. In this case, the educational mode of the program's operation is used.

Taking into account the simplicity of the program, a teacher with basic information preparation can independently develop tests for lessons with the use of remote sensing data.

Generally speaking, the methodology for using aerospace images as a means of visualizing geographic information is largely similar to the method of working with graphical visual aids such as pictures and photographs [5]. Let us note the basic requirements that need to be taken into account when using images as high-quality visual aids:

- the content of aerospace images must be consistent with the content of the material being studied and illustrated at the appropriate time of the lesson;
- the visibility of the images should be used with a reasonable dose of the transmitted information;
- images should be of high quality;
- the remote sensing data collection should be organized on the principle of "from simple to complex" during the study of geography;
- the objects or phenomena depicted in the photographs are to have a geographic binding;
- the information transmitted by the images should not affect the integrity of the lesson;
- the teacher should think over the explanations and comments on the aerospace images.

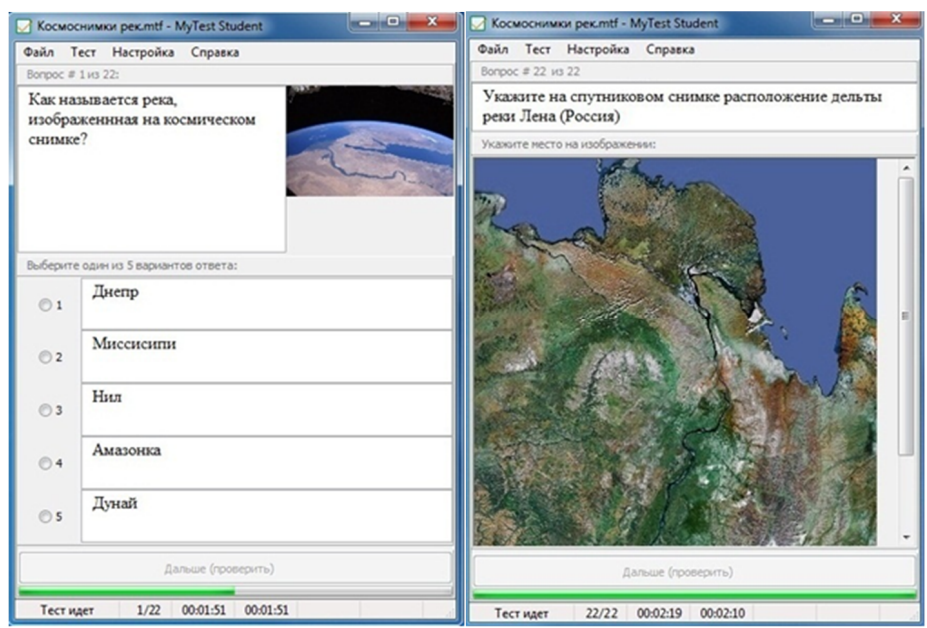

**Fig. 2.** MyTest program windows with different types of questions on the theme "Rivers" with the use of satellite images

## 2.2 Earth remote sensing data as an interactive learning tool (ILT) on geography

Electronic georesources, developed on the basis of satellite images of the Earth, are unfolding new opportunities in the process of geographical education – interactive learning, i.e., learning with feedback, bilateral exchange of information between the subject and the object of the learning process.

Currently, a significant number of interactive learning tools (ILTs) are known in pedagogical practice and geography is one of the leaders in their use [9]. Earth Remote Sensing Data is a unique basis for the creation of interactive geoservices, the skillful use of which allows them to be considered as an ILT with unique educational functions. As an example, we can name the following satellite-based interactive geoservices: Google Earth, Google Maps, NASA World Wind, EINGANA, etc.

By the nature of the transmitted spatially-bound information, all interactive geoservices based on remote sensing data, can be divided into two groups: complex and thematic.

Complex resources (Google Earth, Google Maps, NASA, World Wind, etc.) contain different geoinformation layers (weather conditions, ocean conditions, firewalls, earthquakes, etc.). Thematic georesources (Gismeteo, Meteoweb, Map of Life and etc.) are mono-informational and devoted to certain processes or phenomena (atmospheric processes, traffic flows, migration of animals, etc.). This also includes various spatially-bound social networks.

Wide functionality of such georesources allows them to be used in all the courses of school geography, in various forms of organization of the in-study process, with the involvement of different teaching methods. The most common form of application of ILT based on remote sensing data - classroom, it fits into a traditional lesson, and allows you to organize new types of educational activities. Here are the main types of lessons:

*A. Lesson to learn new teaching material.* The teacher creates in advance labels with information layers that allow you to consistently display static aerospace images, interactive terrain models, photos and video materials on the screen (interactive whiteboard). Unconventional types of lessons in the formation of new knowledge using ILT can include lessons, integrated lessons or research lessons. The basic organizational form of this kind of lessons is working with the class.

*B. Lesson of skills and abilities formation and improvement.* The teacher develops applications in which, using interactive georesourses functions, each student performs individual practical work. The main organizational form is to work with georesources in small groups, although using an interactive whiteboard, the students can work in class collectively.

*C. Lesson of knowledge generalization and systematization.* At the lessons of this type, the teacher offers students creative tasks for laboratory work in the computer class. Independent work on the task, reinforces the cognitive interests of the students, makes their work creative, and in some cases brings it closer to the nature of the research. Unconventional types of lessons of knowledge generalization and systematization can be attributed to students' workshop conferences using data remote sensing and other types of geoinformational technologies.

*D. Lesson of knowledge, skills and abilities control and correction.* The lessons of this type are to control the level of students' assimilation of theoretical material, the formation of skills and abilities; correction of knowledge, accumulated skills and abilities. To this end, the teacher develops control interactive questions based on the use of Google Earth georesourse. An individual or group survey can be as well used in the classroom.

The choice of a lesson form and type depends on its purpose, features of the given class, the studied topic, etc. Of course, extra-curricular work greatly expands the educational potential of the ILT on the basis of remote sensing data. Independent work, optional classes and classes allow students to bring their educational work with interactive aerospace imagery to a completely new level.

### 2.3 Earth remote sensing data as a source of geographic knowledge and skills

The knowledge obtained by analyzing aerospace images, for example, includes such as: spatial position of geographic objects; their morphometric characteristics; qualitative and quantitative indicators; the establishment of causal relationships, patterns, etc. This information is the basis for conducting practical and laboratory classes, as well as students' research work.

The most characteristic feature of the remote sensing data use during the practical work is thorough the images study, with qualitative and quantitative characterization. The knowledge gained during the work with aerospace images within the scope of practical work includes the following: definition of geographical objects size (length, width, perimeter, area, volume) and distances in the area; identification of quantitative and structural indicators (the Earth surface temperature, the spectral brightness of the vegetation, the composition of the forest fund, etc.). It is extremely important that the analysis of aerospace images is to be combined with the use of the whole set of geographic knowledge, including geographic maps, statistical indicators, and field observations.

Laboratory work is a more complex form of practical and research orientation of training realization with the use of remote sensing data. The complication of dealing with aerospace images enhances the influence on the schoolchildren's way of thinking. Tasks performed during laboratory work with the use of geoinformational technologies, promote the students' cognitive activity through the integration of theoretical knowledge and practical skills.

One of the main tasks of laboratory work is to master working skills with aerospace images: their reading and decoding. Of course, laboratory classes should be multileveled, that is, to differ in the complexity of solvable subject and didactic tasks or the method of their conduct.

The complexity of the work to be carried out should gradually increase. At the same time, the design of a series of laboratory works using these remote sensing data should be elaborated taking into account the main areas of students' practical activity.

The Earth image from space for aerial photography is a special means of studying geography, which allows you to create skills about the actual outlines of geographical objects and processes, describe their spatial position, compare different mapping of the terrestrial surface in aerospace images with other sources of geographic information (plan, map, etc.); perform spatial-temporal analysis and so on, thus enriching the world view of the student.

Students can realize the acquired skills by creating a map of the dynamics of the natural environment. The environment is changing: new settlements are emerging and the existing ones disappear; new roads, engineering structures are emerging, new mining areas are being developed; forests are cut down and land use structures are changing. Under the influence of natural and anthropogenic factors shore lines, vegetation and new objects arise. In this regard, the great practical significance is the creation of maps, the main thematic load of which are the boundaries of areas of the territory, exposed to natural or anthropogenic character and, as a consequence, determine the long-term changes in landscapes. The maps of the dynamics are intended to solve tasks connected with the monitoring of the territory and obtaining information about the activity that takes place in the territory.

To create maps of dynamics space images that capture the mapping area received with a certain time interval have been used.

The main purpose of research work organization is the discovery and support of gifted bright students, as well as the development of their intellectual and creative abilities. It is extremely important to choose the right topic for the research. It should be noted that the purpose of the work must be concrete, understandable and accessible. It is necessary that the student in the process of performing the work would realize the practical significance of their research. Taking into account the specificity of the information that is provided by the Earth's observation data, it is necessary to outline the range of issues that could lay the basis for the student's research work.

First and foremost, it is territorial research aimed at solving specific problems in a particular region (for example, monitoring natural and anthropogenic areas). Analyzing time-varying images of the same territory allows students to create dynamic maps that reflect environmental changes: environmental violations, human-induced changes, deforestation, etc.

The second vector of students' research activity is the compilation and refinement of cartographic materials, as well as the registration of the land fund. These works are more applicable and are of great particular interest.

The third direction of the Earth remote sensing data use as a basis for conducting pre-research work is a detailed study of space images in order to identify atypical or unique objects and processes on the Earth's surface.

## 3  Conclusions

1. Earth remote sensing data represent an inexhaustible source of unique information, which opens to the students the door to the world of unknown before. The teacher's task is to enable students to open these doors, since the use of remote sensing data during geography studies has a number of advantages over the traditional teaching materials (e.g. high resolution, high degree of visibility, realism, etc.).
2. The prospects of further scientific research in the use of GIS technologies during the study of geography in profile school and extra-curricular work are regarded as those of top priority.